# Image based Crop Monitoring Technologies in Protected Horticulture: A Review

**Namal Jayasuriya[1*], Yi Guo[2], Wen Hu[3], Oula Ghannoum[1]**

[1]Hawkesbury Institute for the Environment, Hawkesbury Campus, Western Sydney University, Richmond, NSW 2753, Australia

[2]Centre for Research in Mathematics and Data Science, School of Computer, Data and Mathematical Sciences, Western Sydney University, Parramatta, NSW 2150, Australia

[3]School of Computer Science and Engineering, The University of New South Wales, Sydney, NSW 2052, Australia

**\* Correspondence:**
Namal Jayasuriya
N.Jayasuriya@westernsydney.edu.au

## Abstract

Future food security is a major concern of the 21[st] century with the growing global population and climate changes. In addressing these challenges, protected cropping ensures food production year-round and increases crop production per land area by controlling environment conditions. Maintaining the growth and health of crops in these facilities is essential to ensure optimum food production. However, this is a laborious work and is currently done manually. Image-based non-destructive plant phenotyping is an emerging research area that reduces the skilled labour cost while enhancing the monitoring of crop growth, health, and identifying phenotype-genotype relations for plant breeding. With the proliferations of protected infrastructures and targeted plants, different technologies and sensor setups are needed for image-based crop monitoring. Conveyor-type plant-to-sensor systems, bench-top or gantry-based systems are commonly found in research facilities focussing on phenotyping of small, relatively short, or movable model plants. This review examines the literature on crop monitoring and phenotyping platforms in both field and protected facilities and explains different camera technologies and their ability to extract different plant traits. The review highlights the future research directions of image-based monitoring of commercial scale protected crops where crops can be relatively tall or vertically supported under semi controlled environments, which presents new challenges and is rarely covered in the literature.

## 1    Introduction

The global population, projected to reach 9.3 billion by 2050 according to the Department of Economic & Social Affairs of the United Nations (2022), presents a profound challenge for food production. Future food demand is expected to surge by 62%, while the risk of hunger will increase by 30%. Between 2018 and 2022, feeding the world population relies on key commodities including sugar cane, maize, wheat, rice, raw milk of cattle, oil palm fruit, potatoes, soy beans, cassava, and vegetables according to FAOSTAT Production (2023b). All of these productions are involved in high land use, and according to FAOSTAT Land use (2023a), 4.82 billion hectares (37.18%) of land have been utilised for agricultural purposes by 2021. Traditional farming, which is mostly used for those productions, relies on manual labour, seasonal weather, and basic mechanisation. Environmental sustainability and climate change also have an impact on crop production (Foley et al., 2011). Extreme weather events, changing precipitation patterns, and temperature fluctuations challenge crop resilience (Lobell et al., 2011). Addressing the monumental challenge of feeding the world necessitates increasing the production of quality food while minimising the expansion of agricultural lands and minimising the environmental footprint of agriculture, including water usage and agrochemical applications (Godfray et al., 2010; Searchinger et al., 2019).

To reduce environmental footprint, modern practices, such as machinery and agrochemicals, have emerged (Pretty, 2008) and the traditional field agriculture is shifting towards precision and protected agriculture. Precision agriculture uses technologies like remote sensing and automation to optimize resource management (Shafi et al., 2019). Protected agriculture offers control over environmental conditions that enable optimal plant growth, reducing dependence on external weather conditions and better pest management (FAO, 2023; Gruda & Tanny, 2014). However, the operation of advanced protected facilities comes with associated costs, including the high energy expenses incurred in manipulating environmental conditions, the need for pollination of crops that do not support self-pollination, and the demand for skilled labours for crop monitoring and maintenance (Chavan et al., 2022; Samaranayake et al., 2020).

In recent years, the integration of advanced technologies, such as image data acquisition and camera systems, has played a pivotal role in addressing the challenges and opportunities of high-throughput phenotyping (HTP). The purpose of research level HTP, which primarily is conducted in protected environments or small-scale fields, is breeding new crop varieties resilient to climate changes and with improved production. However, the commercial-scale implementation of crop monitoring in protected environments presents infrastructure and crop specific challenges and opportunities (Tian et al., 2022). This article explores the critical nexus of image data acquisition, camera technologies, and the transition to commercial-scale crop monitoring in protected facilities.

In particular, the article will delve into the diversity of protected agriculture infrastructures globally and in Australia, provides an overview of camera technologies for image data acquisition (Table 1), application of camera technologies to extract crop image data (Table 2), dissect the gaps and possibilities associated with the transition from research-oriented HTP to the scalable implementation of these technologies in commercial-scale protected facilities (Table 3) and a framework for selecting imaging platform design to monitor crops in protected horticulture (Table 4). In pursuing this goal, the intention is to shed light on the path forward for sustainable agriculture in an era of increasing food demand and resource constraints.

## 2    Protected Cropping Infrastructures and Their Production Contribution in Australia

Growing crops in protected environments such as traditional greenhouses or more modern plant factories is an alternative method to conventional field agriculture. Protected crops are more productive, efficient, and now becoming popular in both urban and peri-urban areas (Rabbi et al., 2019). According to Australian horticulture statistics 2021/2022, 33 vegetable types are grown in Australia and six of them are also grown under greenhouses, contributing 15.85% to the total production value (Hort Innovation, 2022). Among 29 different fruit types, only blueberries, rubusberries, and strawberries grow under greenhouses as well, contributing 9.15% to the total production value. Fresh flowers and nursery productions were also conducted in greenhouses in addition to conventional field crops, showing a contribution of 13.1% and 16.8%, respectively for flowers and nursery productions. Figure 1 (a) shows the market value contribution from different crop infrastructures for different crop types.

A greenhouse can be defined as an enclosed environment that generates its own microclimate that is conducive for crop growth and development (Bennis et al., 2008; Russo et al., 2014). By controlling parameters such as temperature, $CO_2$ enrichment, humidity, energy heating, and fertigation, farmers can manipulate inputs and outputs. Due to increased demand, advances in greenhouse technology, better utilisation of fertilizer, water, and power consumption, protected cropping has attracted attention across the world (Castilla & Montero, 2008; Torrellas et al., 2012). Furthermore, factors such as the need to reduce carbon footprints, food security, and the growing interest in locally produced food have encouraged governments and farmers to be concerned about protected cropping (Connellan, 2001; Soode et al., 2015). Protected cropping includes polytunnels, glasshouse facilities that enable control of most environmental factors, and fully controlled vertical farming facilities (Chavan et al., 2022). Figure 2 shows examples of different types of protected crop infrastructures.

## 2.1    Polytunnels

Low tech polytunnels contribute 80% to 90% to the protected crop production globally (*Indoor Soilless Farming*, n.d.) and in Australia (*Graeme Smith Consulting - General Industry Information*, n.d.). Low-tech greenhouses typically have transparent or clear cladding materials that are used to cover semi-circular, square, or elongated structures made of bamboo, wood, or metal with polyethylene film (Sivaperuman et al., 2018). Figure 2 (a) shows the exterior and interior of an example low tech protected crop infrastructure. The structure serves to control the water supply and prevention of pests. To have a relative control of heat and humidity, top vent or saw tooth type roof openings and fans are used with vapour spray systems. Soil is used mainly as a growing media in grow bags or pots with automated or semi-automated irrigation systems. Compared to the other greenhouses, it has the benefit of being less expensive. Contrary to high-tech greenhouses, the low-tech greenhouses are not equipped with proper management of environmental characteristics such as temperature, humidity, and $CO_2$ concentration (Awal et al., 2021).

In the Australian context of 2021/2022, 49% of protected production value has come from polytunnels with 14%, 10%, and 24%, respectively have come from vegetable, flower, and nursery productions (Hort Innovation, 2022). For vegetable crops, the majority of cucumber and eggplant productions have come from poly tunnels. Polytunnels have contributed approximately 10% of capsicum and herbs production, and around 5% for both salads and tomatoes (Figure 1 (b)). Polytunnels production by volume (ton) show similar trends to productions value (Figure 1 (c)), with 42% for tomatoes and around 33% for other vegetables compared to glass houses and conventional field crops (Figure 1 (d)).

## 2.2    Glasshouse Facilities

Glasshouses are a type of greenhouses which are covered with glass materials under controlled environments, with varying degrees of control. Some of these greenhouses fall into the category of partially controlled environments, often referred to as medium-tech greenhouses, given the ongoing evolution of protected cropping practices (Chavan et al., 2022). The exterior and interior of an example high-tech semi-controlled glasshouse facility are shown in Figure 2 (b). Within this category, some greenhouses feature automated roof openings, which allow low-cost rudimentary temperature regulation. More advanced structures are equipped with both automated roof opening and sophisticated heating and cooling systems to manage heat with relatively low energy consumption, or only advanced cooling systems like heated liquid flowing systems, fans, chilled air blowers, and air conditioning systems, which can significantly impact energy costs (Samaranayake et al., 2020). In recent developments, smart films, originally used in the construction industry, have found application in greenhouse design to enhance energy efficiency, although many greenhouses continue to employ traditional PVC or glass cladding (Lin et al., 2022). Solar photovoltaic (PV) systems and supplemental lighting, such as LED panels, are other technologies used in large-scale commercial greenhouses to increase crop quality and yields (Marucci et al., 2018). Newly innovated light-shifting films such as LLEAF(Soeriyadi, 2022) are also being trialled to increase crop production without moving to artificial light (Chavan et al., 2020). To optimise plant growth, high-end greenhouses often employ soilless growing mediums, such as rockwool blocks, coupled with precisely calibrated liquid fertilisation techniques, aimed at maximizing crop yields (Ghannoum, 2020). While modern commercial and research-orientated glasshouses are equipped with advanced environmental controls, including temperature, $CO_2$ concentration, humidity, fertigation, and irrigation systems, other aspects such as growth monitoring, crop management, harvesting, and pest/disease control remain predominantly manual, with ongoing research focused on automation (Samaranayake et al., 2020).

Glasshouses account for ~26% of tomato production, while it is around 20% for capsicum and herbs, and 10% for cucumber and salads in 2021/2022 Australia (Figure 1 (b)) (Hort Innovation, 2022). In terms of production value, tomatoes in glasshouses claim 65% of the total tomato market value even if it is 26% in volume, while other vegetable crops show similar trends for market value and production volume (Figure 1 (c)). Figure 1 (d) clearly shows this with 51% of value per ton compared to other crops. Higher rates for value per ton demonstrate the higher quality of food produced in glasshouses compared to other infrastructures.

## 2.3  Vertical Farms

In contrast to greenhouses with various levels of control over environment conditions, fully controlled indoor crops are commonly known as indoor farms or plant factories, representing the pinnacle of controlled environment agriculture. These facilities operate as tightly controlled systems, using 100% artificial lighting to optimise crop production while conserving vital resources such as $CO_2$, water, and temperature. They are mostly inside buildings that are not designed to utilise ambient light directly for plant growth. However, solar panels are used as a renewable energy source to power these plant factories (Rehman, 2022). In particular, fully controlled greenhouses have gained substantial popularity in densely populated countries such as Korea, Japan, and Taiwan (Kozai, 2013). Plant factories designed to grow smaller plants such as lettuce, salad vegies, and microgreens can be seen in two main designs, vertically stacked beds, and towers. Figure 2 (c) shows an example of vertically stacked beds (right) and a vertical tower (left) of salad vegies. LED-based artificial light is typically used to reduce energy consumption and the spectrum range and quality are being optimised for plant growth. All other environment conditions are controlled similarly to semi-automated infrastructures. Hydroponic media are preferentially used together with recycling systems (Mir et al., 2022). In Australia, unlike for other crop infrastructures, 2021/2022 statistics do not show yet production contributions from vertical farms or plant factory facilities according to Hort Innovation (2022).

## 3   Camera Technologies for Crop Image Data Acquisition

Various types of sensors exist where different bands of the electromagnetic spectrum are used, from the ultraviolet range to the thermal range. These sensors used for plant phenotyping can be categorized as RGB imaging, 3D imaging, thermal imaging, fluorescence imaging, hyperspectral imaging, and tomographic imaging. Each category targets some specific traits of plants and some of these traits can be identified using multiple sensors with different accuracy levels. Figure 3 visualises where different imaging techniques operate along the electromagnetic spectrum as well as the sensitivity of traits in the electromagnetic spectrum. Table 1 summarises the traits that are under the sensitivity of different sensors or cameras targeting shoot phenotyping while grouping them as morphological, biophysical, or biochemical traits.

### 3.1   RGB Cameras

The most popular and cost-effective method for measuring plant or organ morphological traits, biomass, and plant growth is RGB imaging, which is also known as visible light imaging (Yang et al., 2014). However, RGB imaging cannot reveal physiological information. The RGB imaging equipment, which are normal cameras, capture the RGB range of the electromagnetic spectrum (400-750 nm) that is identifiable to the human naked eye. Chlorophyll absorbs photons from the blue and red regions of the spectrum and reflects most of the photons in the green region. The green colour of leaves occurs as a result of this reflection, and the change of chlorophyll content can be easily identified using RGB imaging. In addition to measuring colour change, leaf disease identification and counting of leaves and fruits are also possible. A visible camera, which is relatively fixed to a plant, can also be used to measure structural traits such as image-based projected biomass, leaf area, and growth dynamics (L. Li et al., 2014a).

### 3.2   3D scanners

3D imaging is more accurate for obtaining spatial and volumetric traits such as leaf area, plant volume, height, and projected biomass estimation compared to normal RGB imaging. Numerous 3D imaging techniques, including laser scanning (Paulus et al., 2014), Light Detection and Ranging (LiDAR) scanning (Hosoi et al., 2011), structured light (Omasa et al., 2003), stereo vision (Rovira-Más et al., 2005), structure-from-motion (Paturkar et al., 2020b), space carving (Golbach et al., 2016), and others, have been developed over the past 15 years as a result of technological advancement. The user should choose the imaging technique according to the application because each imaging technique has its own advantages and disadvantages. Structure from motion (SfM) technique has performed well in the outdoor scenario, as in Paturkar, Gupta and Bailey, (2020a). The computation time was a significant limiting factor because SfM needs a lot of images to produce a 3D model and the right number of images must be chosen to generate a precise 3D model according to in-depth analytical study by Paturkar et al,. (2021).

### 3.3   Thermal Imaging

Infrared radiation can be seen in thermal imaging, which shows an object as the temperature of its surface. The most frequently used wavelengths for thermal imaging are 3-5 µm or 7-14 µm, and the sensitive spectral range of thermal cameras is 3-14 µm (L. Li et al., 2014a). Thermal imaging measures leaf surface temperature to study water relations and stomatal conductance of plants. A major determinant of leaf temperature is the rate of evaporation or transpiration. In response to various biotic and/or abiotic stresses, plants often experience decreased rates of photosynthesis and transpiration (Bellvert et al., 2014; Cohen et al., 2015). Remote sensing of leaf temperature using thermal imaging can be a reliable method of identifying these changes. Infrared sensors are also used to measure the surface temperature. Although thermal cameras cost more and require more handling than infrared sensors, they have several advantages over infrared sensors, including better spatial resolution and more accurate measurements under a variety of environmental conditions (L. Li et al., 2014b). Typically, the canopy temperature and background temperature are included in the thermal images of the crops. Eliminating background noise from thermal images is a crucial issue, particularly for temperatures correlated to edge pixels that contain both canopy and background (Prashar & Jones, 2016). Therefore, an accurate calibration is needed for surface temperature measurements.

### 3.4  Fluorescence Imaging

Fluorescence is the term used for light that is produced by the chlorophyll complex when some shorter wavelengths of radiation are absorbed during photosynthesis. When the chloroplasts are exposed to blue or actinic light, some of the light that was absorbed by the chlorophyll is re-emitted. The plant's capacity to metabolise the harvested light affects how much re-emission light there is compared to the amount of irradiation. Fluorescence imaging is the imaging of these fluorescence signals, and it typically makes use of charge-coupled device (CCD) cameras that are sensitive to fluorescence signals. Fluorescence signals are produced when samples are illuminated with visible or ultraviolet (UV) light using pulsed lasers, pulsed flashlight lamps, or LEDs (Gorbe & Calatayud, 2012). Chlorophyll fluorescence imaging is a potent tool for resolving the spatial heterogeneity of leaf photosynthetic performance and has been applied in a variety of plant physiology applications, including early identification of stress symptoms by pathogen invasion or herbicide application (Bürling et al., 2010; Rolfe & Scholes, 2010). Most fluorescence imaging studies are restricted to model crop seedlings or single leaves. To address the use of large-scale phenotyping and create a standardised method for processing fluorescence images, robustness, reproducibility, and data analysis software are required (L. Li et al., 2014a). Furthermore, field phenotyping applications might be constrained by the power requirements of fluorescence imaging (for instance, by using short-wave laser stimulation).

### 3.5  Hyperspectral and multispectral

A typical hyperspectral camera covers 400 nm to 1000 nm spectral bands, which include almost all the VIS (400-700 nm) and the NIR (700-1200 nm) ranges. Compared to the other sensors, a hyperspectral sensor gives a number of bandwidth ranges as its output, and each bandwidth can be a few nanometres depending on the sensor. The output is called a "hypercube" which looks like a number of stacked images of the scene, where each image refers to one bandwidth. Hyperspectral imaging is used for both morphological and biochemical feature extraction. Similar to an RGB camera, plant height, leaf area, leaf shape, and biomass can be estimated as morphological features using a hyperspectral camera that is relatively fixed relative to the object or with the use of 3D reconstruction algorithms with a number of images. Disease identification, sugar content, water content, chlorophyll content, and micronutrients such as nitrogen, potassium, and phosphorus content are the biochemical features that can be investigated using a hyperspectral camera.

Numerous hyperspectral remote sensing experiments have shown that the usual spectral sign of macronutrient stress in plants is a decrease in red-edge slope (Pacumbaba Jr & Beyl, 2011). The red-edge slope is assumed to be associated with reduced chlorophyll levels, and its inflection point lies in the 700–709 nm spectral region (Carter & Knapp, 2001). The spectral area between 780 and 787 nm, which is in the near infrared section of the radiometric spectrum, is used as the foundation for the computation of normalised difference vegetation index (NDVI) in a number of hyperspectral remote sensing investigations (Haboudane et al., 2004; Manley, 2014). The accumulation of above-ground biomass, chlorophyll content, and plant N levels have all been shown to be substantially associated with NDVI (Boelman et al., 2003; Jones et al., 2007). However, processing the data from the sensor (sample size is in gigabits) might be difficult and costly due to the complexity of data, high volume, and computational resources required (Mahlein et al., 2018). For quantitative analysis, the hyperspectral reflectance is constantly dependent on the lighting conditions and is found challenging under ambient light, which is varying. Imaging at night with artificial light could be an effective substitute (Nguyen et al., 2020). The same plants may produce different reflectance patterns depending on whether hyperspectral imaging data are acquired during the day or at night. Further research is needed on the usage of active light sources and the issue of whether imaging at night or during the day provides the greatest window of opportunity for identifying and assessing agricultural plant responses and biochemical status.

## 3.6 Tomographic Imaging

Tomographic imaging includes Nuclear Magnetic Resonance Imaging (MRI), Positron Emission Tomography (PET), and X-ray imaging. Seeds (Melkus et al., 2011), complete root systems grown in soil (Moradi et al., 2010), and entire plants (Van As & Van Duynhoven, 2013) can all be studied using MRI by obtaining 3D datasets. Additionally, MRI can be used to estimate water diffusion and transport, describe 3D representations of water distribution, and quantify plant water content without the need for an invasive procedure (Windt et al., 2006). It can also be used to find labelled molecules (Melkus et al., 2011).

PET is also a nuclear imaging technique that creates a 3D representation of a functional process. Through a radionuclide that emits positrons, it indirectly detects pairs of gamma rays. It is capable of non-invasively imaging the distribution of labelled compounds like $^{11}$C (Jahnke et al., 2009), $^{13}$N (Kiyomiya et al., 2001), or $^{52}$Fe (Tsukamoto et al., 2009). PET can also image the transport of $^{11}$C-labelled photo assimilates during photosynthesis. This imaging mode can dissect transport domains in plant organs and measure transport velocity and lateral loss (Bühler et al., 2011). Combined with MRI, it can provide structural and functional traits and analyse the transport of water and labelled compounds.

X-ray computed tomography creates tomographic images of particular areas of the scanned object using computer-processed X-rays. It can also create a 3D image of an object's interior using a series of 2D radiographic images taken around a single axis of rotation. This method can measure the architecture of the root system (Flavel et al., 2012) and can provide volumetric data for various structures with different densities, such as soil structural heterogeneity and plant structures. It has been used on a variety of species, including barley, maize, wheat, chickpea, Arabidopsis, and other root system architecture systems (L. Li et al., 2014b). Cost and scanning time are the drawbacks of X-ray CT.

## 4    Applications of Camera Technologies for Crop Monitoring

### 4.1    Conveyor Systems

Plants in the pots are delivered into an image chamber made up of several cameras in a conveyor-style plant to sensor architecture. For data gathering, the chamber is often a dark room with cameras mounted on the top and sides, or the plants are rotated. The automated door eliminates interference from ambient light, and halogen bulbs provide illumination. A typical conveyor-type HTP "Scanalyzer 3D" developed by LemnaTec GmbH (Aachen, Germany), which also covers the modified versions, has been accepted by certain international organisations and states the following (Yang et al., 2020). A variant known as "Plant Accelerator" was produced by the Australian Plant Phenomics Facility (APPF). Chickpea salt tolerance and crop nutrient insufficiency have both been studied effectively using it (Atieno et al., 2017; Neilson et al., 2015) . It has various imaging units for RGB, fluorescence, NIR, and hyperspectral imaging, which are dispersed and have autonomous operations, and it can accommodate 2,400 plants. Using only two hyperspectral cameras in their hyperspectral imaging chamber, Bruning *et al.*, (2019) recently assessed the spatial distribution and concentrations of water and nitrogen levels in wheat. In another case, the conveyor belt setup of a smart house was used to investigate the impact of fungi and zinc (Zn) on tomato cultivation (Brien et al., 2020).

Scanalyzer 3D, housed at the University of Nebraska-Greenhouse Lincoln's Innovation Center, enables the phenotyping of 672 plants up to 2.5 metres tall by gathering RGB, NIR, IR, fluorescence, and hyperspectral data from the top and side views of the plants. A revolving elevator is available in each imaging room to view the plant in 360º (Das Choudhury et al., 2018). To reach the target weight, three watering stations are there with scales. Ge et al., (2016) used RGB and hyperspectral imaging spaces to examine the dynamics of maize's growth and water utilization. For the first time in vivo, Pandey *et al.*, (2017) measured plant nutrient concentration and water content using the Scanalyzer 3D hyperspectral imaging chamber. In order to find genetic connections, Miao *et al.*, (2020) segmented hyperspectral pictures of sorghum and maize at the organ level.

A growing chamber from Conviron (Winnipeg, Canada) and an imaging station (LemnaTec Scanalyzer) are part of the bellwether phenotyping platform (Fahlgren et al., 2015). RFIDs are used as identifiers of plant post to correlate image data with information. 1,140 plants may be transported to the VIS, fluorescence and NIR imaging stations using a 180 m long conveyor belt. The four components of this conveyor system can operate alone or together, improving experimentation and research versatility. Due to the microclimatic variability that is introduced when plants are moved from a greenhouse to an imaging room, Purdue University created cyclic conveyor belts where plants can complete their entire growth cycle (Ma et al., 2019).  Huazhong University of Science and Technology and Huazhong Agricultural University (Wuhan, China) created a high-throughput rice phenotyping facility (HRPF) that can track 15 agronomic parameters of 1,920 rice plants (Yang et al., 2014). This HRPF measured the drought response in rice (Duan et al., 2018). Conveyor HTP systems need greater flexibility and are expensive. With automatic weighing and watering, the commercially available PlantScreen system moved trays each containing 20 pots, on conveyor belts from the growing area to a dark acclimation chamber and RGB and chlorophyll fluorescence imaging chambers (Awlia et al., 2016). Czedik-Eysenberg et al. (2018) created PhenoBox, a versatile open-hardware and open-source phenotyping device to extract shoot features to offer less expensive phenotyping alternatives. PhenoBox was frequently used to research infection of model grass, salt stress, and observation of other crops with additional sensors, such as hyperspectral sensors (Czedik-Eysenberg et al., 2018). To identify morphometric traits (aboveground biomass, plant morphology, and plant colour) that may be related to yield, CropDesign (Belgium) developed TraitMill, which consists of proprietary bioinformatics tools, a high-throughput gene engineering system, plant transformation techniques, and a HTP platform (Reuzeau et al., 2006). However, the specifics of TraitMill, the structure of the trial, and the results were all confidential.

Plant silk, which can be found on plants such as corn, is difficult to identify and measure because of its unique characteristics. This was overcome by Phenoarch (INRA, Montpellier, France) using a HTP device with two imaging cabins to track the development of corn ears and silks. The plants are first rotated at a steady pace using RGB cameras to identify their ear coordinates. For continuing to collect high-resolution photographs of the silk, the robotic arm places the camera 30 cm from the ear. This makes it possible to monitor the daily development of the ear and silk of hundreds of plants.

**4.2    Benchtop systems, Gantry systems, and fixed camera systems**

Arabidopsis and other small plants are the focus of benchtop phenotyping. By keeping the plants steady and moving the imaging system above the plants, it adheres to the sensor to plant approach. The broad spread, quick life cycle, and compact genome of *Arabidopsis thaliana* make it the perfect model plant for tabletop HTP. 1440 *Arabidopsis thaliana* plants can be accommodated on the Phenovator platform (Flood et al., 2016). Its eight-position filter wheel for its monochrome camera measures projected leaf area (PLA) at eight wavelengths and photosynthesis. The Phenoscope imaging device comprises of a digital camera and can accommodate 735 pots (Tisne et al., 2013). The pots are continually rotated, ensuring that the tested plant is exposed to the same external circumstances. This reduces microenvironmental variance at the individual plant level and creates a space with a high degree of homogeneity.

For top views of tiny plants, the LemnaTec Scanalyzer HTS features a robotic arm with VIS, NIR, and fluorescence cameras. It has been used to research the time dependent water stress in *Arabidopsis thaliana* (Acosta-Gamboa et al., 2022). Research on molecular plant physiology is being constrained by the identification of mutant phenotypes (Fraas & Lüthen, 2015). Chang *et al.*, (2020) gathered growth photos of 350 *Arabidopsis thaliana* plants and analysed the subtle morphological impacts of various radiation doses over the plant's growing season to learn more about the phenotypic traits of dose effects and information about dynamic growth behaviour.

Gantry-based systems, which are larger in size than benchtop-style systems, are targeted at somewhat larger plants than Arabidopsis, including wheat and rice. A platform called Glyph is made up of four bridge-like structures. On the platform, a gantry consists of drip irrigation tools and a camera travelling between pairs of rows. It has been used successfully to predict soybeans' ability to withstand dry fields (Peirone et al., 2018). Shakoor et al. have employed a fixed camera platform using low-cost Raspberry Pi computers equipped with standard eight-megapixel camera modules. These cameras are strategically positioned as an array at the roof level with top-down views in a greenhouse, allowing for automated and high-resolution image capturing. The plants (a relatively small grass variety) were grown in pots and spread on the floor of the greenhouse. This setup enables 3D reconstruction of individual plants and monitoring of temporal changes in plant phenotypes over time.

The Field Scanalyzer, developed by Rothamsted Research, is a rail-based gantry system for field phenotyping that measures 125 metres in length, 31.5 metres in width, and 36 metres in height. It is fitted with a 300 kg sensor array that includes a visible camera, a 3D laser scanner, a thermal infrared camera, a visible to near-infrared (VNIR) hyperspectral camera, a four-channel amplified radiometer, chlorophyll fluorescence sensor, and $CO_2$ sensor. For one plot, data from all sensors were acquired in about 7 minutes. This platform can be used, for instance, to calculate the timing of wheat ear emergence and flowering, to create a complete description of crop canopy growth across all life stages (Sadeghi-Tehran et al., 2017). Guo *et al.*, (2018) created a high-throughput crop phenotyping platform (Crop3D) to measure 3D plant/leaf structures and leaf temperature using an integrated LiDAR sensor, RGB camera, thermal camera, and hyperspectral imager in a mobile gantry system. However, there are two difficulties that need to be addressed: (1) The issues posed by varying ambient light are partially alleviated by midnight imaging or the adoption of an on-field reflectance reference (Virlet et al., 2016). (2) Effectively analysing terabytes of hyperspectral pictures and laser-scanned images, as well as interpreting temporal series of thermal, multispectral, or hyperspectral-based features in changing contexts, are difficult tasks. Furthermore, the cost of building and maintaining such massive rail-based systems over such a small region is high. LeasyScan is a gantry system run on a rail system for experimental field crops targeting HTP. The laser scanner is equipped with the gantry to 3D scan plants targeting leaf area and structure (Vadez et al., 2015).

### 4.3 Mobile robots

The Vivo robotic system is designed for the autonomous measurement of maize and sorghum leaf characteristics. It consists of a four degrees-of-freedom (DOF) manipulator, a time-of-flight (TOF) camera, and a gripper integrated with an optical fibre cable and thermistor (Atefi et al., 2019). More specifically, the gripper can monitor VNIR spectral reflectance and temperature, and the TOF camera serves as the vision system. However, the capture speed and success rate of the gripper (78%for maize and 48%for sorghum) could be improved.

Spi-see is a camera system designed for tall vertically supported capsicum crops in a glasshouse (Polder et al., 2009; van der Heijden et al., 2012). It consists of four vertically stacked RGB and ToF camera modules to capture depth information and 3D reconstruct the view. Artificial light, a blue screen behind the plants to filter the background of the scene and QR codes to identify plants are used. These 3D data have been used to extract structural features such as plant height, area of a leaf, and angle of a leaf.

Phenobot is another robot that targets tomato fruits in a glasshouse environment (Fonteijn et al., 2021). It uses a camera module consisting of depth and RGB modules by Intel to extract 3D structural features and they have been used with both RGB and depth data to segment tomato fruits.

In the open field, the majority of imaging platforms are vulnerable to changing environmental factors such as light intensity. The BreedVision device simply eliminates ambient light and conducts imaging within a portable dark chamber to tackle this issue (Busemeyer et al., 2013). BreedVision can non-destructively measure a variety of plant traits, including plant height, tiller density, grain yield, moisture content, leaf colour, lodging, and dry biomass, at an operating speed of 0.5 metres per second because it is equipped with multiple sensors including a ToF, laser distance sensors, hyperspectral imaging, RGB sensors, and light curtain imaging. Surprisingly, without the impact of indoor or field illumination circumstances, light curtain imaging can be used to measure plant height, plant architecture, and biomass when equipped with a pair of linear light barriers (one emitter and the other one as receiver) (Busemeyer et al., 2013; Fanourakis et al., 2014).

Bonirob is a tractor type robot vehicle which has a sensor setup that goes over the plant to scan the top view of a crop (Ruckelshausen et al., 2009). It consists of RGB, lidar and hyperspectral sensors targeting leaf area, NDVI index, and colour index of maize and wheat. PocketPlant3D is a mobile phone based system which measures canopy leaf traits(Confalonieri et al., 2017). This is an easy to use and affordable solution for farmers, but it is limited to a few traits and does not have a robust model to face complex conditions in the crop.

Vinobot (Shafiekhani et al., 2017) and Robotanist (Mueller-Sim et al., 2017) are phenotyping robots designed for field crops. Both of them consist of either stereo vision or lidar sensors with RGB cameras to extract structural measures of grass varieties such as maize, wheat, or sorghum. The main targeted features are leaf area, plant height, and different colour indices. Separation of individual plants for plant level features is not feasible due to overlap and occlusion problems, and the results of Robatanist have to be validated.

### 4.4 UAVs/Drones

Unmanned Aerial Vehicles (UAVs) such as drones offer a versatile platform that may quickly collect data over vast areas and possibly produce photographs with a high spatial resolution up to 1 mm per pixel. Computer science techniques such as Deep Learning are capable of processing millions of remote sensing images accurately and quickly (Bauer et al., 2019). In order to monitor drought stress response, evaluate nutritional status and growth, identify QTLs, and detect diseases, remote sensing has been employed extensively (Maes & Steppe, 2019; Wang et al., 2019). The high spatial and temporal resolution, canopy colour, and texture features obtained by UAV platforms make phenotyping jobs easier, since better image quality and quantity provide detailed data for feature mining and analysis (Yue et al., 2019). As a result, high-resolution UAV photography has been used for a variety of phenotyping tasks, including weed detection, wheat ear identification, leaf area index estimate, and rapeseed seedling performance evaluation (Yao *et al.*, 2017; Madec *et al.*, 2019; Zhao *et al.*, 2018). Additionally, the determination of the best resolution has received a lot of attention from scholars. For instance, a recent study found that HTP in the field was possible with an ideal resolution of 0.3 mm (Madec et al., 2019). UAV-based phenotyping opens new opportunities in field research due to its combination of high spatial and temporal resolution (Burkart et al., 2018).

## 5 Gaps and Research Directions for large scale crop monitoring in Protected facilities

### 5.1 Image data acquisition for commercial indoor crops

In Section 2, the current status of commercial crops in Australia was discussed (summarised in Figure1) and it shows that the majority of crops are still cultivated in the field except a few berry varieties, capsicum, tomato, eggplant and cucumber. Statistics of vegetable production show that production quality and value per ton are higher in glasshouse facilities and polytunnels compared to conventional field crops. However, only selected crops are grown under greenhouse conditions. Those crops can be divided into two categories according to their size and nature: 1) relatively short plants (salad, herb, berry, flower, and nursery productions); 2) taller and vertically supported plants (tomatoes, capsicums, eggplants).

There are two main types of data acquisition systems: 1) plant to sensor systems which move the plant to a camera system such as imaging chambers with cover systems; 2) sensor to plant systems which move the camera systems closer to the plants such as gantry or mobile robots. A visual summary of imaging data acquisition platforms which are categorised according to infrastructure and usage is shown in Figure 4.

Category 1 plants are relatively short plants that are grown in glasshouses, poly tunnels, or poly houses in Australia. However, they can also be grown in vertical farms as stacked beds in rack structures or as vertical towers (Mir et al., 2022). Most of the discussed phenotyping systems have used small model plants, such as Arabidopsis or grass varieties, which are relatively short or are not supported vertically. Hence, most of these phenotyping systems can be considered suitable to monitor category 1 commercial crops. Plant to sensor conveyor systems are not practical even for small crops such as micro greens, because moving each plant or plot of plants to an imaging chamber requires a substantial mechanical system. Gantry systems to move a single camera unit to plants or many fixed camera units to capture a large area following sensor to plant architecture are suitable for this category. Since plants are not too high, top-down and top-angled views are good enough for 3D reconstruction and other visual feature extractions. However, the gantry system will be cost effective if expensive camera types are targeted to be used in the camera unit than deploying many expensive camera units to cover a large area. Vertically stacked beds in vertical farms or typical greenhouses are the suitable infrastructures for deploying gantry-based systems. A fixed camera setup would be good if it is a tower type vertical farm and towers can be moved to the fixed camera setup. Otherwise, a mobile robot will be suitable if towers are organised as rows. Mobile robots are also suitable for indoor facilities when plants or plots are organised as rows, allowing a mobile robot to move along plant lines to scan and image them. Future research can be conducted to study the cost effectiveness of gantry based single camera unit versus fixed multiple camera units in greenhouse facilities and fixed camera system versus mobile robot for tower type plant factories.

Category 2 includes relatively tall vertically supported plants that grow primarily in glass houses and polytunnels. They are basically supported with strings attached to the ceiling level of the greenhouses and sensor units also deployed above the crop to monitor the environment conditions (Tian et al., 2022) as the information provided in Section 2.2. These obstacles make it difficult for gantry-based systems to be used for category 2 crops. For crops such as capsicum, cucumber, and eggplant in this category which have mature fruits in the mid and lower levels of the plants; diseases and infections are possible to start appearing at the lower canopy level where the dying leaves are (J. Li et al., 2021). Hence, gantry-based systems are not good for this category as they are not good for feature extraction at the mid or bottom level canopy even if they are somehow deployed at ceiling level. Fixed camera setups also face similar problems with the need of having many camera units to cover a large area. There is no way of using conveyor systems even for research purposes as plants cannot be moved away. Mobile robots have been shown as an option to acquire data for this crop category. Vertically movable cameras or fixed stacked up camera setups can provide the necessary degree of freedom to capture close-up images at any level of crop height. They can also be extended to harvest (Fonteijn et al., 2021) or pollinate (Dingley et al., 2022) plants as research is going on. In addition to the accessibility of different height levels of the crop, there are few other challenges that can be faced. These crops eventually get very dense, and crop management practices such as pruning are used to control the number of branches and dense. Separate identification of plants and organs also becomes harder when plants are very close to each other due to overlap and occlusion(Tian et al., 2022). There is a gap in studies that uses mobile robots to automate the monitoring of large-scale vertically supported tall plants in green house facilities.

Drones can be less practical for both categories in greenhouses unless they have to carry only a tiny light weight camera, because a big drone with a heavy payload generates wind, which can vigorously shake plant leaves. This disturbs the 3D reconstruction for structural feature extraction or image stitching. However, future research can be conducted to test the usability of tiny drones or to solve the problems that occur due to the wind generated by relatively large drones. Mobile applications developed for plant trait extraction are also not practical as they need someone to carry them to all plants and trigger the app manually. It is suitable to monitor a few randomly selected plants manually. The suitability of different image data acquisition systems for different crop categories in greenhouse facilities is summarised in Table 3 and a framework for selecting appropriate image data acquisition design is shown in Table 4 considering arrangement of plants, plant type, and light condition.

**5.2    Challenges for camera technologies to use at different greenhouse infrastructures.**

Regarding the usability of different camera technologies in greenhouses, the primary problems are light condition and occlusions. Hyperspectral and fluorescence cameras strictly require controlled light conditions (Yang et al., 2020) and have been used under controlled light conditions in the literature discussed in Sections 3 and 4. Images taken with RGB cameras also get disturbed with varying light conditions, especially due to noises such as bright spots and shadowing, which can become dominant and vanish visual features. Because of varying light conditions, foreground-background separation using image processing techniques also becomes difficult (Razmjooy & Estrela, 2019). The mobile dark chamber is an option for imaging under ambient light in greenhouses that people have tried for field crops (Busemeyer et al., 2013) and it may work for gantry systems which target the top-down view of category 1 plants. For mobile robots designed to image vertically supported tall plants, it is too complex to cover the ambient light as the camera looks at plants from side or angle views. As an alternative, people have used those cameras with artificial light at night time (Nguyen et al., 2020). Fully controlled high tech glasshouses and plant factories do not face issues due to varying light conditions as the light is fully artificial and controlled. However, the filtered spectrum optimised for plant growth in plant factories also disturbs image processing based foreground-background separation (Tian et al., 2022). Future research is needed to study the usability of cameras under ambient light conditions, filtered spectrum, and the practicability of mobile dark chambers with gantry systems.

CT scanning systems such as MRI and PET are mainly used for imaging the root systems of plants that cannot be captured with other camera technologies, as discussed in Section 3.6. MRI and PET machines are large enclosed systems which use strong magnetic fields and objects need to move into the system to be scanned. Hence, this is not practical to monitor large scale crops which needs sensor to plant approach. X-ray systems are radiation systems that are used to scan objects and they are harmful to living beings. Hence, they are used as enclosed environments and not suitable for green house crops which are open spaces where humans are involved in activities. Future research is needed to study new technologies that can be used to scan root systems in open spaces or adaptation of CT and radiation technologies to use as sensor to plant systems.

**6    Conclusions**

The review discussed the statistics of Australian agriculture production in 2021/2022 in Section 2.2 and reveals that protected agriculture produces quality production that leads to more value while utilising less space. The literature has identified energy, pollination, and labour cost as the main challenges for protected agriculture. Diverse of camera technologies that can be used for different purposes were studied next. After that, the review studied the crop image data acquisition platforms in both fields and protected facilities which can be used to reduce the skilled labour requirement for crop monitoring. In the discussion and gap identification, the adaptability of existing phenotyping system designs to monitor large scale crops considering two general crop types: relatively small plants and vertically supported taller plants were discussed. The discussion shows that gantry-based systems are more suitable for monitoring relatively short crop types while showing the suitability of mobile robots to monitor vertically supported tall commercial crops. The discussion on the challenges of camera technologies to monitor commercial indoor crops, the difficulty of using hyperspectral and fluorescence in green houses with uncontrolled light conditions, and disturbance to feature identification using RGB images due to varying light conditions and occlusions were shown. The problems of using CT and X-ray scanners to scan the root system are also highlighted. The review proposes mobile robots in green house facilities to monitor the growth of vertically supported commercial crops targeting image-based features extracted at different levels of plant height, gantry based mobile dark chambers following field mobile dark chambers, research on using hyperspectral and fluorescence cameras under ambient light conditions, and adapting existing technologies to monitor root systems under open environment conditions as future research directions.

# 7    Figures

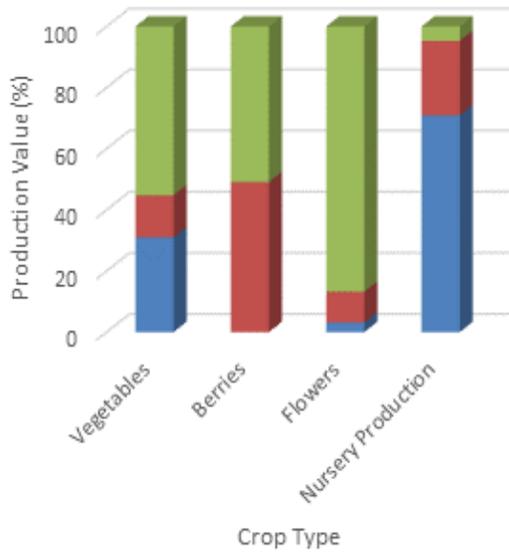
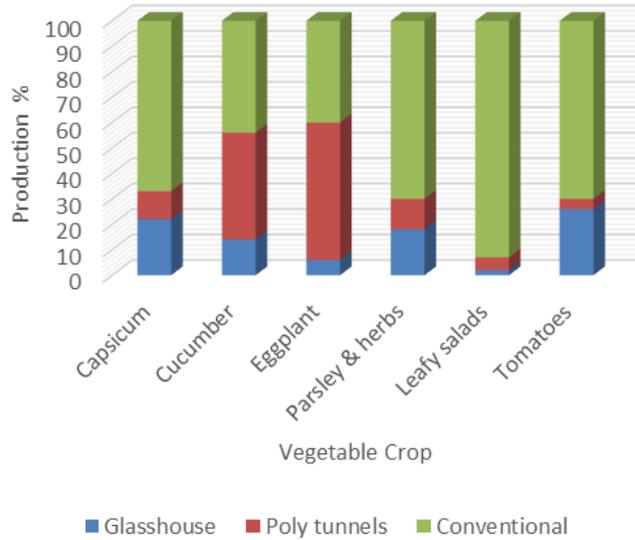

(a) (b)

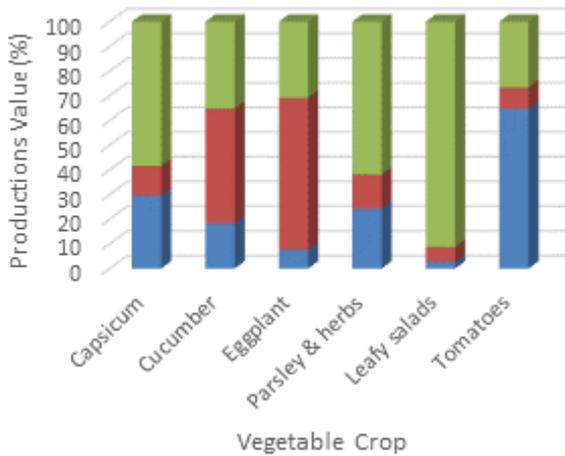
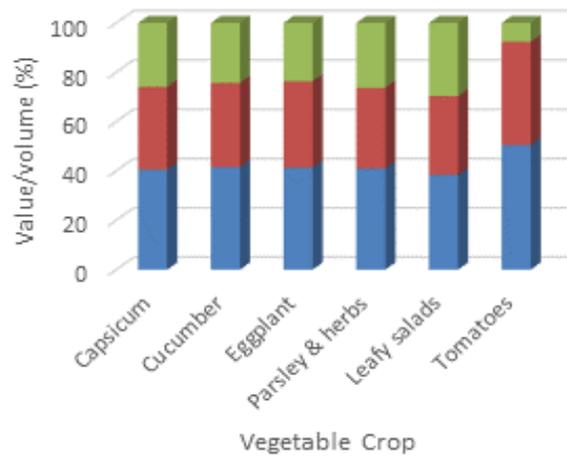

(c) (d)

Figure 1: Overview of distribution of crops and their market contribution in different infrastructures for crops which are currently commercially grown under protected facilities. Adapted from Australian Horticulture Statistics Handbook 2021/2022 (Hort Innovation, 2022).

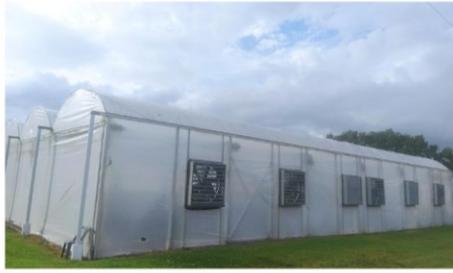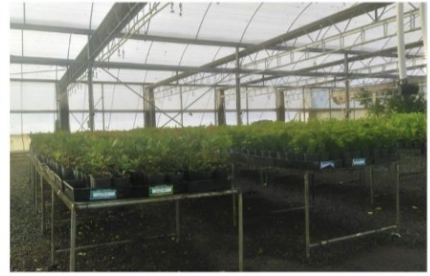

(a) Low-tech PC

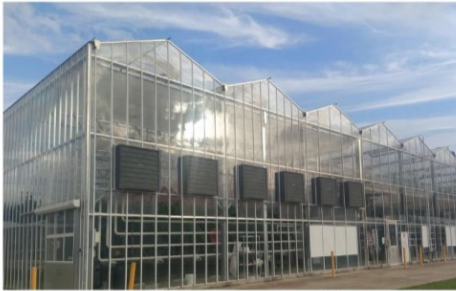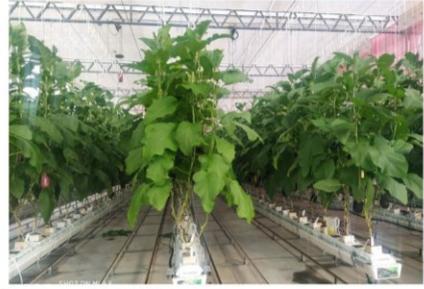

(b) High-tech semi-controlled PC

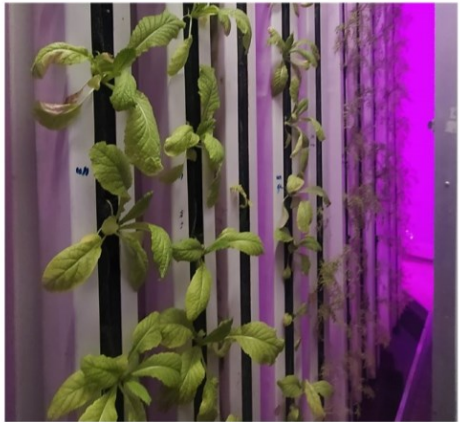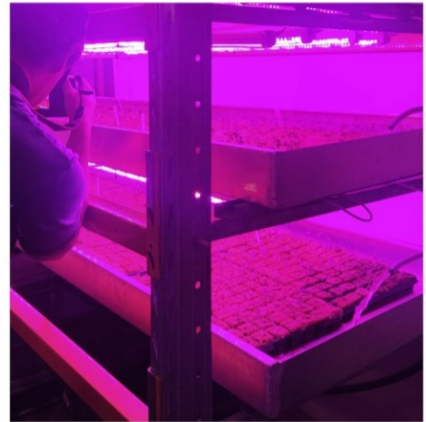

(c) High-tech fully-controlled PC

Figure 2: Diversity of protected crop infrastructures according to technology level and design.

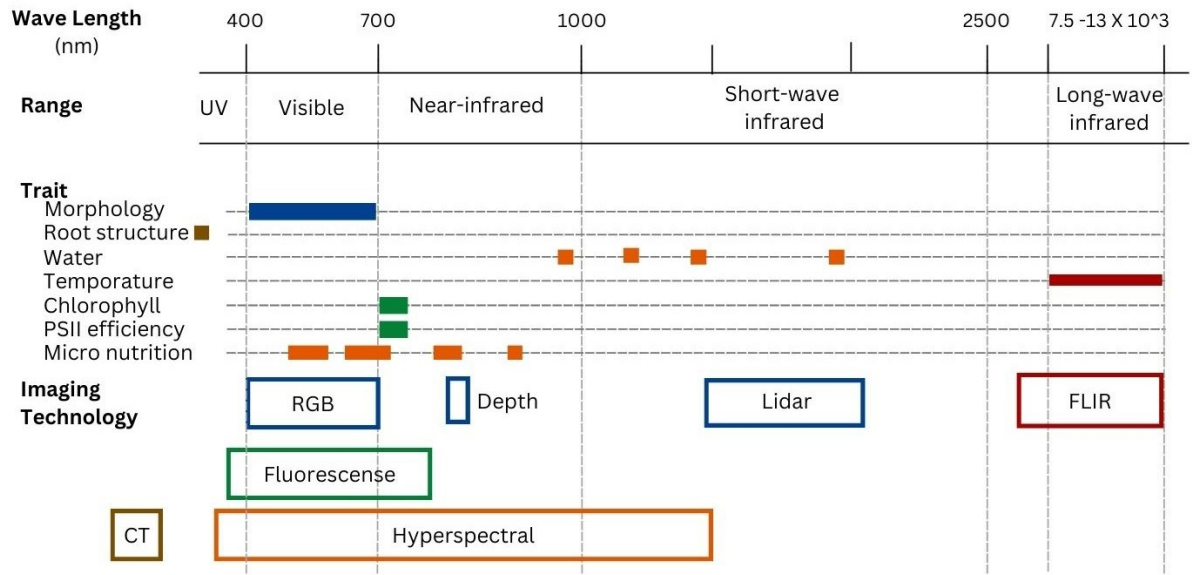

Figure 3: Electromagnetic Spectrum and Imaging Techniques for Plant Phenotyping. The figure illustrates the location of different imaging techniques in the electromagnetic spectrum and the sensitivity of plant traits to electromagnetic radiation. The top section shows how the spectrum is divided into different regions and their wavelengths. The middle section indicates the sensitive ranges of plant phenotyping traits, while the lower section shows the sensitive ranges of different camera technologies. The colours in the lower section indicate which camera technologies are most effective for capturing different plant traits.

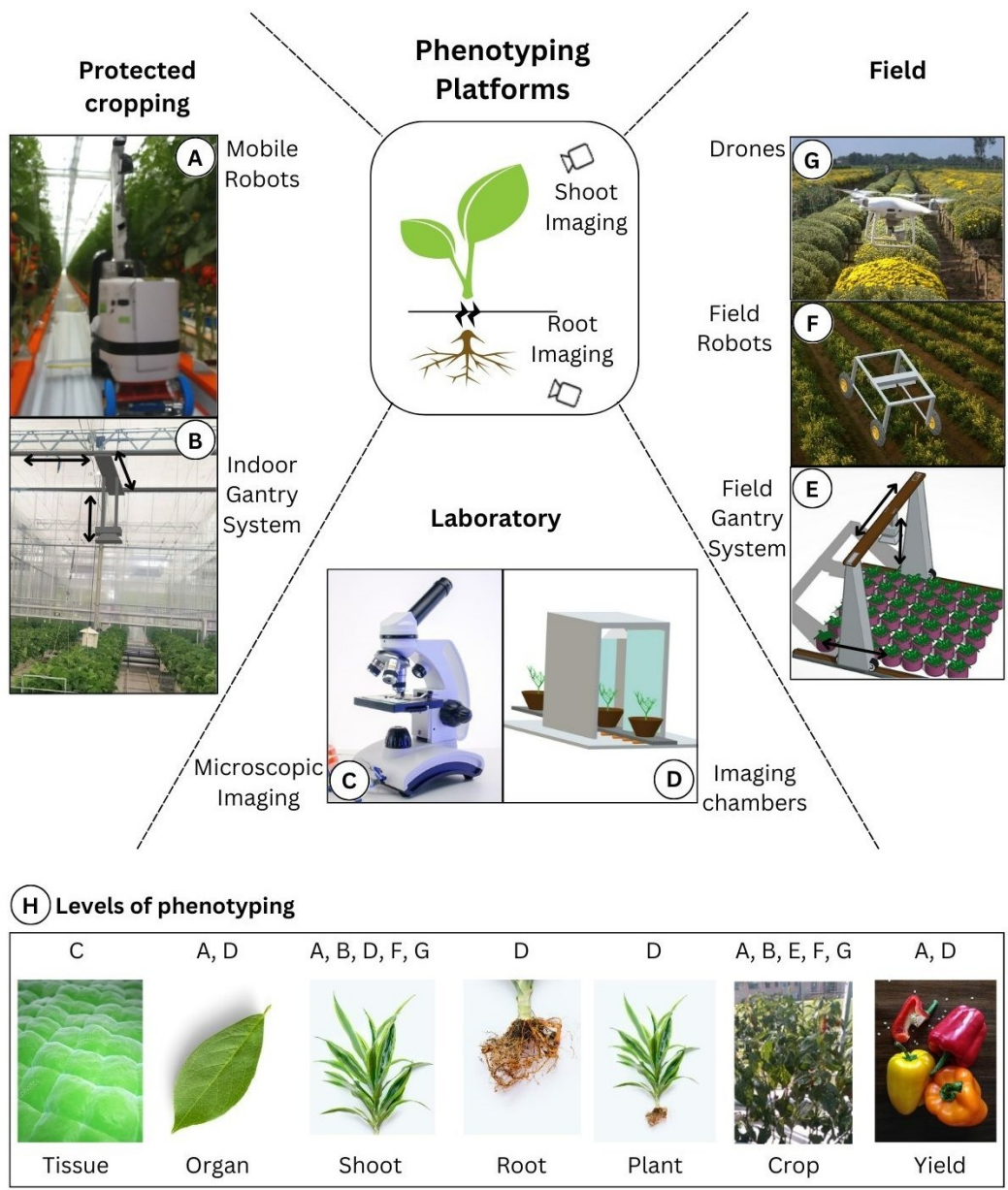

Figure 4: Categorization of image-based plant phenotyping platforms considering their infrastructure and their applications. Panel A shows a robotic platform capable of targeting shoot and organ level imaging in protected crop environments. Panel B illustrates a model of gantry-based system that can be used with top-down cameras for shoot phenotyping. The two arrows show the degree of freedom of the camera system. Panel C highlights the use of microscopic imaging for organ and tissue level phenotyping in laboratory settings. Panel D presents a model of phenotyping chamber capable of more advanced levels of phenotyping under fully controlled environments, where plants are transported using conveyor belts. Hyperspectral and fluorescence imaging, as well as radiation-based techniques like X-ray and CT, can be performed in this architecture. Panel E visualises a model of field gantry system that can be used to monitor plants with a top-down view. The arrows show the degree of freedom of the camera unit. A modelled tractor in a field crop equipped cameras mounted on top of it, providing a downward-facing view is shown in Panel F. Panel G depicts drones or light aircraft, targeting crop level imaging. Panel H shows the different levels of phenotyping from tissue

to crop level. The letters on top of the images in Panel H show the mapping of each level of phenotyping with different phenotyping approaches.

# 8  Tables

Table 1: Sensitivity of cameras or sensors to morphological, biophysical, or biochemical traits in plant shoot phenotyping.

| Camera / Sensor | Morphological traits | Biophysical traits | Biochemical traits |
| --- | --- | --- | --- |
| RGB camera | plant height, canopy cover, biomass (with a fixed camera) | stay green, diseases | chlorophyll content |
| Stereo camera | plant height, canopy cover, biomass | | |
| Lidar camera | plant height, canopy cover, biomass | | |
| Fluorescence camera | biomass | light use efficiency, disease | |
| Multispectral camera | plant height, canopy cover, biomass (with a fixed camera) | diseases | chlorophyll content, nitrogen content, water content |
| Hyperspectral camera | plant height, canopy cover, biomass (with a fixed camera) | diseases | chlorophyll content, nitrogen content, water content, sugar content |
| Thermal camera | canopy cover | diseases, stomatal conductance, surface temperature | water content |
| Photosynthesis sensor | biomass | light use efficiency, diseases | |
| Leaf area sensor | leaf area | | |

Table 2: Existing sensor platforms related to phenotyping of protected cropping or beneficial for protected cropping.

| Platform | Environment / System type | Sensors | Crop | Traits | Limitations | Reference |
|---|---|---|---|---|---|---|
| Lemma Tec Scanalyzer | Glasshouse / Conveyor type | RGB, NIR, Fluorescence, Hyperspectral | Barley, Sorghum, Maize, Wheat, Chick pea, movable plants | Biomass, plant height, width, compactness, leaf water content, nitrogen content, salt stress | Designed for small movable plants. Imaging under controlled conditions. | (Atieno et al., 2017; Bruning et al., 2019; Miao et al., 2020; Neilson et al., 2015; Neumann et al., 2015; Yang et al., 2014) |
| Bellwether | Glasshouse / Conveyor type | RGB, NIR, Fluorescence | Seteria | Plant height, biomass, water-use, water content | Designed for small movable plants. Imaging under controlled conditions. | (Fahlgren et al., 2015) |
| Bellwether | Glasshouse / Conveyor type (circular conveyor ) | Hyperspectral | Maize | PLA, NDVI, perimeter, major axis length, minor axis length, eccentricity | Designed for small movable plants. Imaging under controlled conditions. | (Ma et al., 2019) |
| HRPF | Glasshouse / Conveyor type | RGB, CT | Rice | Drought stress, tiller number | Designed for small movable plants. Less flexibility. | (Duan et al., 2018; Yang et al., 2014) |
| PlantScreen | Glasshouse / Conveyor type | FLUO, RGB | Indoor crops | Photosynthetic efficiency, colour and shape-based features | Suits for movable plants. Multiple controlled imaging cabinets. | (Awlia et al., 2016) |
| PhenoBox | Glasshouse / Conveyor type | RGB | Movable indoor plants | Salt stress, disease infection | Labour intensive for large screening. Suits for movable plants | (Czedik-Eysenberg et al., 2018) |
| TraitMill | Glasshouse / Conveyor type | RGB | Monocots | Colour, height, surface | Proprietary, movable plants | (Reuzeau et al., 2006) |
| Phenovator | Greenhouse / Benchtop type | Monochrome camera with 8 wavelengths | Arabidopsis thaliana | Projected leaf area, photosynthesis | Designed for small plants under controlled environments. | (Flood et al., 2016) |
| Phenoscope | Greenhouse / Benchtop type | RGB | Arabidopsis thaliana | Rosette size, expansion rate, evaporation | Designed for small plants under controlled environments which can be moved easily with the pot. | (Tisne et al., 2013) |
| Glyph | Greenhouse / Benchtop type | RGB | soybean | Water use efficiency, drought stress | Designed for small or medium plants. High cost for exact setup. | (Peirone et al., 2018) |
| Field Scanalyzer | Field crop / Gantry based | VIS, NIR, FLUO, Hyperspectral, Multispectral, 3D Laser, IR, | Field or indoor crops. | Height, colour, nutrient content, water content, 3d structure, temperature | Costly; Costly to extend to a large crop area; variable ambient light | (Sadeghi-Tehran et al., 2017; Virlet et al., 2016) |
| Crop 3D | Field crop / Gantry based | LiDAR, RGB, Thermal, Hyperspectral | Grass varieties | plant/leaf structures, leaf temperature, biological measures for genomic analysis | Costly to extend for large crop area | (Guo et al., 2018), |
| LeasyScan | Field crop / Gantry based | Laser scanner | pearl millet | Leaf area, canopy structure | Only structural measures, Costly to extend for large crop area | (Vadez et al., 2015) |
| BreedVision | Field crop / Mobile dark chamber | RGB, Hyperspectral, 3D ToF, Depth | Movable plants | Height, tiler density, grain yield, moisture content, colour, dry biomass | Restricted by wet soil and weather conditions (rainy and strong breeze) | (Busemeyer et al., 2013) |
| Vinobot | Field crop / Mobile robot | Stereo vision, Lidar | Maize, sorghum | Plant height, leaf area index | Designed for outdoor crops. Separation of plants | (Shafiekhani et al., 2017) |

| | | | | | is not clear for individual plant phenotyping. | |
|---|---|---|---|---|---|---|
| BoniRob | Field crop / Mobile robot | RGB, Lidar, Range camera, Hyperspectral | Maize, wheat | Color, leaf area index, NDVI | Suitable for field phenotyping from the top of plants. | (Ruckelshausen et al., 2009) |
| Robotanist | Field crop / Mobile robot | RGB, Stereo vision, Range camera, Lidar | Corn, sorghum | stalk geometry, leaf erectness, leaf necrosis, GRVI, and spectral indices. | Phenotyping results are needed to be validated with this robot. | (Mueller-Sim et al., 2017) |
| Vivo Robot | Greenhouse / Mobile robot | ToF camera, Spectrum sensor, Thermistor | Maize, sorghum | Leaf traits: temperature, water content, chlorophyll, potassium. | Leaf level features only, moderate accuracy. | (Atefi et al., 2019) |
| Spi-see | Greenhouse / Mobile robot | RGB, Stereo vision, Distance sensor (ToF) | Capsicum | Height, single leaf area | Targets only QTL mapping | (Polder et al., 2009; van der Heijden et al., 2012) |
| Phenobot | Greenhouse / Mobile robot | RGB, Stereo vision | Tomato | Tomato fruit detection and fruit growth features: size, colour | Targets only fruit | (Fonteijn et al., 2021) |
| Drones or UAVs | Field crop / Flying cameras | VIS, NIR. Multispectral, Hyperspectral | Field crops | Colour, water content, nutrient level, canopy cover | Low resolution, strict operating, and local flight laws should be noted to ensure flight safety. | (Hung et al., 2014; Madec et al., 2019; Maes & Steppe, 2019; Yue et al., 2019; Zhao et al., 2018) |
| PocketPlant 3D, | Field for Indoor / Mobile | Smartphone camera | Maize | Canopy and leaf traits | Lack of robust models to face complex conditions in the field | (Confalonieri et al., 2017) |

Table 3: Usability of the studied image data acquisition systems to monitor large scale crops in protected facilities.

| Imaging System | Number of Systems Studied (Field / Greenhouse) | Usability in large scale PC monitoring | Drawbacks in large scale PC monitoring |
| --- | --- | --- | --- |
| **Fixed cameras** | 0 / 1 | Suitable only for small crops such as micro greens which occupy less space. Suitable for vertical farming systems where the vertical towers are movable. | Many cameras are needed to be deployed to cover a large-scale crop. Not good to exact traits of taller plants which needs better side view. Not good for vertically supported crops which have more obstacles on top of plants. |
| **Conveyor Systems** | 0 / 7 | The plants should be movable to imaging system | Not practical. |
| **Gantry / Bench top Systems** | 3 / 3 | Suitable for relatively short plants. Suitable for vertical farming systems with stacked layers. | Not good to exact traits of taller plants which needs better side view. Not good for vertically supported crops which have more obstacles on top of plants. |
| **Mobile robots / vehicles** | 4 / 3 | Suitable for crops where plants are arranged as lines. Can be used for vertically supported taller plants and small plants as well. | Less practical for crops arranged as plots. |
| **Drones / UAVs** | 5 / 0 | Tiny drones are suitable for crops which have more space above the crop. | Drones generate more wind with their size and load increases. It creates fast movements of plant leaves creating newer problems for applications that depend on 3D reconstruction. |
| **Mobile Device Apps** | 1 | Good only for checking a few random plants | Not practical for large crops as someone needs to carry it and trigger it to collect data. |

Table 4: Ccategorisation of crop organisation, light condition and plant type, in protected facilities as a framework of selecting imaging platforms for crop monitoring in protected horticulture.

Conditions for hyperspectral, multispectral or florescence imaging: 0 – direct, 1 – with additional lights, 2 – dark chamber with artificial light, 3 – night time with artificial light

|  |  | Single layer (typical green house) | | Multilayer (vertical farming) | |
|---|---|---|---|---|---|
|  |  | Rows | Plots | Towers | Racks |
| Ambient light (varying) | Short/small plants | Mobile robot[3], Gantry system[2], Fixed camera system[3] | Gantry system[2], Fixed camera system[3] | Mobile robot[3], Gantry system[2], Fixed camera system[3] |  |
|  | Tall / Vertically supported plants | Mobile robot[3] |  |  |  |
| Artificial light (fully controlled) | Short/small plants | Mobile robot[0,1], Gantry system[1] | Gantry system[1], Fixed camera system[0] | Mobile robot[0,1], Gantry system[0,1], Fixed camera system[0,1] | Gantry system[0,1], Fixed camera system[0] |
|  | Vertically supported plants | Mobile robot[1] |  |  |  |

## 9 Conflict of Interest

The authors declare that they have no conflict of interest.

## 10 Author Contributions

W.H., Y.G., N.J. and O.G. designed the research. N.J. prepared the manuscript with contributions from all co-authors.

## 11 Funding

Research support for this study was provided by a project entitled "IoT for Indoor Farming" funded by the Future Food Systems CRC (FFSCRC), UNSW and WBS Technologies awarded to Prof Wen Hu and Prof Oula Ghannoum. NJ was supported by a scholarship funded by the FFSCRC and Western Sydney University.

## 12  Acknowledgments

Western Sydney University and FFS-CRC for their financial support, and National Vegetable Protected Cropping Centre (NVPCC) at Western Sydney University are acknowledged for providing Protected Cropping infrastructure for this study.

## 13  Data Availability Statement

Not applicable